\documentclass[aps,
prb,
reprint,
preprintnumbers,
superscriptaddress,
amsfonts,
amssymb,
amsmath
]{revtex4-2}
\usepackage{bm,latexsym,mathrsfs,enumerate,bigints}
\usepackage{upgreek}
\usepackage[table,x11names]{xcolor}
\usepackage[breaklinks=true,
unicode=true,
urlcolor = RoyalBlue4,
colorlinks = true,
citecolor = RoyalBlue4,
linkcolor = RoyalBlue4
]{hyperref}
\usepackage{graphicx}

\usepackage{mathtools}
\usepackage[cal=boondoxo,frak=boondox,scr=rsfs]{mathalfa}

\renewcommand{\vec}[1]{\bm{#1}}
\DeclareMathOperator{\arctanh}{arctanh}
\DeclareMathOperator{\sign}{sign}

\begin{document}

\title{Boundary conditions for the N\'{e}el order parameter in a chiral antiferromagnetic slab}

\date{January, 22, 2021}

\author{Oleksandr~V.~Pylypovskyi}
\email{o.pylypovskyi@hzdr.de}
\affiliation{Helmholtz-Zentrum Dresden-Rossendorf e.V., Institute of Ion Beam Physics and Materials Research, 01328 Dresden, Germany}
\affiliation{Kyiv Academic University, Kyiv 03142, Ukraine}

\author{Artem V. Tomilo}
\email{tomilo.art.2018@knu.ua}
\affiliation{Taras Shevchenko National University of Kyiv, 01601 Kyiv, Ukraine}

\author{Denis D. Sheka}
\email{sheka@knu.ua}
\affiliation{Taras Shevchenko National University of Kyiv, 01601 Kyiv, Ukraine}

\author{J\"{u}rgen Fassbender}
\email{j.fassbender@hzdr.de}
\affiliation{Helmholtz-Zentrum Dresden-Rossendorf e.V., Institute of Ion Beam Physics and Materials Research, 01328 Dresden, Germany}

\author{Denys Makarov}
\email{d.makarov@hzdr.de}
\affiliation{Helmholtz-Zentrum Dresden-Rossendorf e.V., Institute of Ion Beam Physics and Materials Research, 01328 Dresden, Germany}

\begin{abstract}
Understanding of the interaction of antiferromagnetic solitons including domain walls and skyrmions with boundaries of chiral antiferromagnetic slabs is important for the design of prospective antiferromagnetic spintronic devices. Here, we derive the transition from spin lattice to micromagnetic nonlinear $\sigma$-model with the corresponding boundary conditions for a chiral cubic G-type antiferromagnet and analyze the impact of the slab boundaries and antisymmetric exchange (Dzyaloshinskii--Moriya interaction) on the vector order parameter. We apply this model to evaluate modifications of antiferromagnetic domain walls and skyrmions upon interaction with boundaries for different strengths of the antisymmetric exchange. Due to the presence of the antisymmetric exchange, both types of antiferromagnetic solitons become broader when approaching the boundary and transform to a mixed Bloch--N\'{e}el structure. Both textures feel the boundary at the distance of about 5 magnetic lengths. In this respect, our model provides design rules for antiferromagnetic racetracks, which can support bulk-like properties of solitons.
\end{abstract}

\maketitle

\section{Introduction}

The requirement for high storage densities and operation speed of devices stimulates the development of antiferromagnetic (AFM) spintronics and spin-orbitronics~\cite{Jungwirth16,Gomonay17,Baltz18,Yan20}. The envisioned devices rely on AFM textures moving in spatially confined channels~\cite{Barker16,Gomonay16a,Jin16a,Xia17,Shen18b,Sanchez-Tejerina20} similarly to ferromagnetic racetracks~\cite{Parkin08}.
One of the most efficient ways to control their dynamics are spin-orbit staggered torques, which require specific symmetry of antiferromagnets rendering them chiral with Dzyaloshinskii--Moriya interaction (DMI)~\cite{Zelezny14,Zhang14,Manchon19,Wornle20}. In this respect, the technological progress in design and optimization of AFM racetracks requires a fundamental understanding of the interaction of magnetic solitons with boundaries of a chiral AFM slab. 

Similarly to ferromagnets~\cite{Jiang15,Mueller16}, sample boundaries in antiferromagnets usually act as an injector of solitons~\cite{Khoshlahni19} and alter the shape of a spatially confined domain wall in the media with patterned surfaces~\cite{Hedrich20}. The inhomogeneous DMI of the surface type leads to the surface twist of the order parameter in a two-dimensional (2D) antiferromagnet~\cite{Lund20}. If a homogeneous DMI is present in addition, an enhanced surface magnetization accompanies the deviation of the N\'{e}el vector from the collinear state~\cite{Lund20}. 

To understand properties of the ground state and AFM solitons in confined geometries, it is necessary to make a proper transition from the Heisenberg spin lattice to the micromagnetic model~\cite{Ivanov05a}. The behavior of AFM lattices can be described using two alternative approaches. Historically, the first one was proposed in the seminal works of Louis N\'{e}el (for review we refer the reader to~\cite{Barbara19}). This framework utilizes a representation of antiferromagnetically coupled ferromagnetic sublattices $\vec{M}_1$ and $\vec{M}_2$ with $|\vec{M}_{1,2}| = M_\textsc{s}$ being the saturation magnetization~\cite{Turov01en}. The corresponding boundary conditions are derived for each of the sublattices~\cite{Stamps84,Stamps87,Ghader19,Ghader19a} also allowing to take into account demagnetizing fields~\cite{Stamps87,Stamps84}. Within the second approach, the equations of motion for $\vec{M}_{1,2}$ are rewritten in terms of dimensionless vectors of N\'{e}el $\vec{n} = \left(\vec{M}_1 - \vec{M}_2\right)/ (2M_\textsc{s}) $ and ferromagnetism $\vec{m} = \left(\vec{M}_1 + \vec{M}_2\right)/ (2M_\textsc{s} $). For many practical cases, the relation $|\vec{m}| \ll |\vec{n}| \approx 1$ is fulfilled and it is possible to exclude the vector of ferromagnetism as being a slave variable. Then, the resulting model of the AFM contains a single vector order parameter instead of two \cite{Baryakhtar79}. The same can be obtained using purely symmetrical approach~\cite{Andreev80}. Importantly, although they are different in methodology, both approaches lead to the same soliton equations~\cite{Mikeska80}.

The micromagnetic formulation of the behavior of $\vec{n}$ in one dimension (1D) can be done by splitting the spin lattice into dimers~\cite{Ivanov95e,Mikeska04,Tveten16,Pylypovskyi20}. A straightforward procedure shows that the continuum Lagrangian contains the topological term proportional to $\vec{m}\cdot \partial_x \vec{n}$, which determines differences between the quantum integer and half-integer spin chains~\cite{Affleck89}, e.g., the Haldane gap. It originates from the choice of dimer pairs: the Hamiltonian of a spin chain is not invariant with respect to the sublattice exchange. The latter also leads to the intrinsic magnetization of 1D textures~\cite{Tveten16}. For the case of 2D AFM lattices, the terms $\vec{m}\cdot \partial_i \vec{n}$ are not as important as for spin chains~\cite{Ivanov95e}. However, the dimerization in 2D bipartite lattices is ambiguous since it can be performed along one of two independent direction ($i=x$ or $i=y$) and may lead to spurious effects due to the choice of spin pairs~\cite{Papanicolaou95a,Papanicolaou97}. This issue can be overcome by splitting e.~g., of the square lattice into tetramers~\cite{Komineas98,Komineas20b}. While still there are only two order parameters, the master N\'{e}el vector (director) $\vec{n}$ and the slave ferromagnetism vector $\vec{m}$, this procedure requires two additional auxiliary fields behaving as spatial derivatives of the N\'{e}el vector. This transition allows to build a nonlinear $\sigma$-model of a chiral 2D antiferromagnet preserving the spatial invariance within the lattice plane~\cite{Komineas98,Komineas20b}.
 
Although 1D and 2D cases are fairly well understood, there is no rigorous transition from a spin lattice description to a three-dimensional (3D) micromagnetic model. This leads to a gap in our understanding of the impact of boundary conditions on the ground state and antiferromagnetic solitons in spatially confined chiral AFMs. The boundary of a sample is an additional source of the symmetry break in AFM lattices. Thus, a unit antiferromagnetic cell should be properly chosen to correctly determine the boundary conditions for the order parameters.

\begin{figure*}
\includegraphics[width=\linewidth]{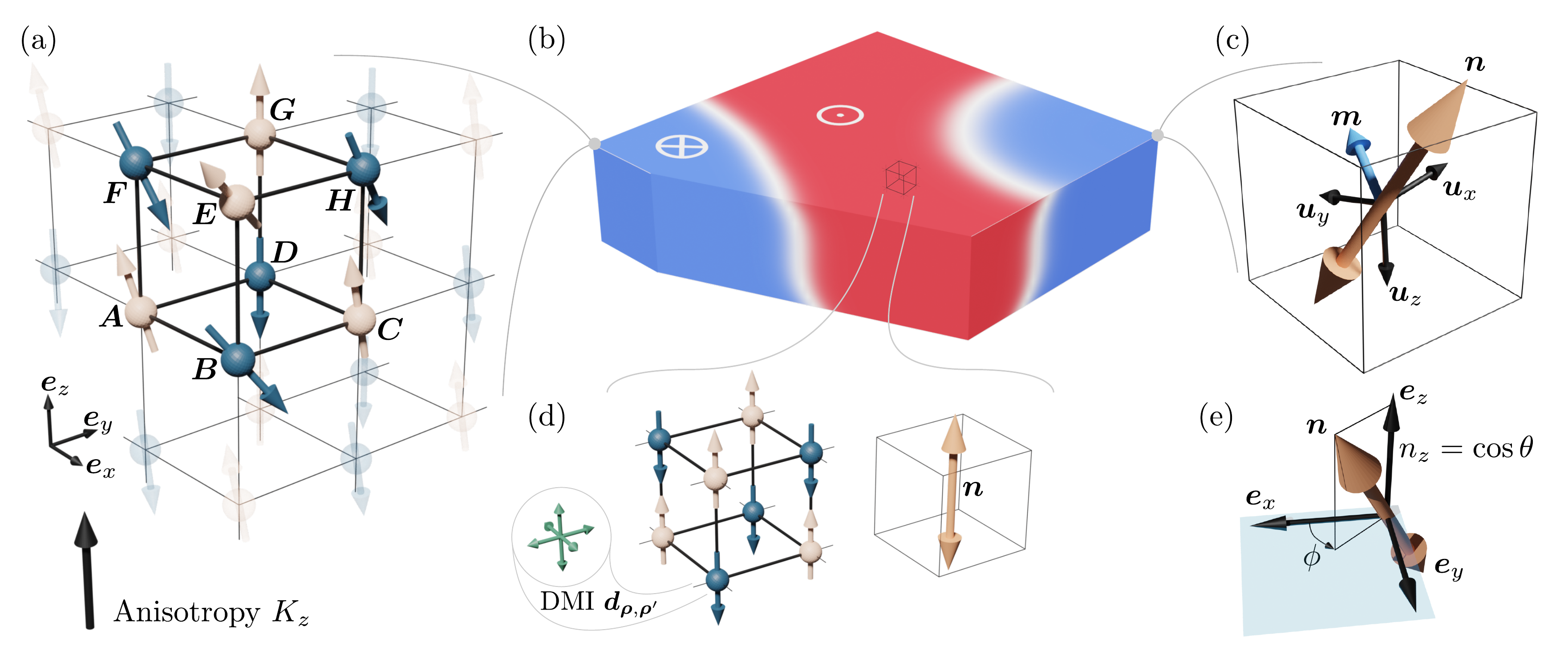}
\caption{\textbf{Chiral G-type antiferromagnet.} Simple cubic crystal lattice with antiferromagnetically coupled spins {(a)} at the edge of a slab~{(b)}. The highlighted octamer $\vec{\rho} = \{i,j,k\}$ contains 8 spins, labeled as $\vec{A},\ldots,\vec{H}$ (index $\vec{\rho}$ is omitted). Less neighbors at the edge as compared to the bulk results in a tilt of spin moments from the anisotropy direction due to the exchange and DMI. {(b)}~AFM slab is schematically colored to show two chiral domain walls, tilted at the edges of the slab. {(c)}~The continuum counterpart of the octamer {(a)} with the N\'{e}el vector $\vec{n}$, magnetization vector $\vec{m}$ and auxiliary vector fields $\vec{u}_{x,y,z}$. {(d)} The spin octamer in the bulk and its continuum counterpart posses no tilt from the anisotropy direction. The DMI vector is shown for the bulk DMI type. {(e)} Parametrization of the N\'{e}el vector $\vec{n}$ in the local spherical reference frame with the polar and azimuthal angles $\theta$ and $\phi$, respectively.}
\label{fig:intro}
\end{figure*}

Here, we rigorously derive a transition from the classical spin-lattice Heisenberg Hamiltonian to the nonlinear $\sigma$-model for a 3D chiral AFM slab with a simple cubic lattice by means of the octamerization process. We avoid spurious parity-breaking effects and show the influence of boundary conditions and DMI on the ground state in the laterally confined sample. We apply this model for two types of magnetic solitons, namely, a translational AFM domain wall and skyrmion, and demonstrate their modification due to the sample boundaries. These textures possess twists in the order parameter as well as the deformation of their shape near the surface, which is analyzed quantitatively. 

This paper is structured as follows. The spin-lattice Hamiltonian and derivation of the corresponding nonlinear $\sigma$-model with boundary conditions is described in Sec.~\ref{sec:sigma}. In Sec.~\ref{sec:ground}, this micromagnetic model is applied for the description of the ground state of a chiral uniaxial AFM slab. Properties of the domain wall and skyrmion in a confined geometry are described in Sec.~\ref{sec:dw} and~\ref{sec:skyrmion}, respectively. The main results of the paper are summarized in~\ref{sec:conclusions}. Further details describing the transition from the spin lattice to the continuum description are provided in Appendices~\ref{app:spinlattice} and~\ref{app:surface}. Appendices~\ref{app:slasi} and~\ref{app:spinflop} contain the description of spin-lattice simulations. The models for the domain wall and skyrmion are discussed in Appendices~\ref{app:dw} and~\ref{app:sk}, respectively.

\section{Results and Discussion}

\subsection{Nonlinear $\sigma$-model}
\label{sec:sigma}

We consider a chiral, uniaxial G-type antiferromagnet with a simple cubic lattice with the lattice constant $a_0$. It can be characterized by the following Hamiltonian including exchange, anisotropy and DMI terms
\begin{subequations}\label{eq:model-g}
\begin{equation}\label{eq:hamiltonian}
\begin{split}
\mathscr{H} & = \dfrac{\mathscr{J}S^2}{2} \sum_{\vec{\rho},\vec{\rho}'} \vec{\mu}_{\vec{\rho}}\cdot \vec{\mu}_{\vec{\rho}'} - \dfrac{\mathscr{K}S^2}{2} \sum_{\vec{\rho}} (\mu_{\vec{\rho}}^z)^2 \\
& + \dfrac{S^2}{2} \sum_{\vec{\rho},\vec{\rho}'} \vec{d}_{\vec{\rho},\vec{\rho}'} \cdot \left[ \vec{\mu}_{\vec{\rho}}\times \vec{\mu}_{\vec{\rho}'}  \right].
\end{split}
\end{equation}
Here, $\mathscr{J} > 0$ is the exchange integral, $S$ is the spin length, $\vec{\mu}$ is the unit magnetic moment in the lattice site enumerated by vector index $\vec{\rho} = \{i,j,k\}$ with $\vec{\rho}'$ running over all nearest neighbors, $\mathscr{K}$ is the constant of uniaxial anisotropy and $\vec{d}_{\vec{\rho},\vec{\rho}'} = -\vec{d}_{\vec{\rho}',\vec{\rho}}$ is the DMI vector\cite{Dzyaloshinsky58,Moriya60a,Yang15}. The dynamics of the magnetic moments is described by the Landau--Lifshitz equation~\cite{Landau35}
\begin{equation}\label{eq:LL}
\partial_t \vec{\mu}_{\vec{\rho}} = \dfrac{1}{\hslash S} \vec{\mu}_{\vec{\rho}} \times \dfrac{\partial \mathscr{H}}{\partial \vec{\mu}_{\vec{\rho}}}
\end{equation}
\end{subequations}
with $\hslash$ being the Planck constant. 
The characteristic length scale for the spin lattice is given by the magnetic length $\ell = a_0 \sqrt{\mathscr{J}/|\mathscr{K}|}$. 

To develop the continuum counterpart of~\eqref{eq:model-g}, we divide the 3D spin lattice by groups of spin octamers with spins being labeled from $\vec{A}_{\vec{\rho}}$ to $\vec{H_{\vec{\rho}}}$ within the given octamer, see Fig.~\ref{fig:intro}(a) and Appendix~\ref{app:spinlattice}. This approach is a 3D counterpart of the tetramerization scheme used for the case of a 2D AFM \cite{Komineas98,Komineas20b}. The magnetic state of each spin octamer is described by vectors of the total magnetic moment $\vec{m}_{\vec{\rho}}$, the N\'{e}el vector (staggered magnetic moment) $\vec{n}_{\vec{\rho}}$ and auxiliary fields $\vec{u}_{\vec{\rho}\,i}$, $\overline{\vec{u}}_{\vec{\rho}\,i}$, $i = x,y,z$, see Fig.~\ref{fig:intro}(c). In the following, we perform the analysis in a long-wave approximation (i.e. spatial and temporal variations of the vector fields are slow) using $\epsilon = \sqrt{|\mathscr{K}|/\mathscr{J}}=  a_0/\ell \to 0$ as a scaling parameter. This implies that while $\mathscr{K}\sim \epsilon^2$, the DMI $\vec{d}_{\vec{\rho},\vec{\rho}'}$ and time derivatives are of the order of $\epsilon$. The ground state is given by $|\vec{n}_{\vec{\rho}}| = 1$. This suggests that the length of the magnetization and auxiliary vectors is of the order of $\epsilon$, see details in Appendix~\ref{app:spinlattice}. Then, the relation between the continuum counterparts of the order parameters and auxiliary fields is given by the
linear expansion of Eq.~\eqref{eq:LL} with respect to $\epsilon$
\begin{equation}\label{eq:order-params}
\vec{m} = -\dfrac{1}{12} \dfrac{\hslash}{\mathscr{J}S} \vec{n}\times \partial_t \vec{n},\quad
\vec{u}_i  = - \dfrac{a_0}{2} \partial_i \vec{n}, \quad \overline{\vec{u}}_i  = 0.
\end{equation}
Thus, three auxiliary vectors are determined by the spatial derivatives of the N\'{e}el vector and other are zero due to the lattice symmetry. This is similar to the case of a 2D antiferromagnet~\cite{Papanicolaou95a,Papanicolaou97,Komineas20b}, where the unit cell consists of four neighboring spins arranged in square.

In the main text, we focus on the chiral AFM slabs with the DMI of the bulk type which is commonly found in AFM crystals~\cite{Dzyaloshinsky58,Moriya60a}. For completeness, the case of the surface DMI is discussed in Appendix~\ref{app:surface}. For the case of bulk DMI, the DMI vector is $\vec{d}_{\vec{\rho},\vec{\rho}'} = d \vec{e}_{\vec{\rho},\vec{\rho}'}$ with $\vec{e}_{\vec{\rho},\vec{\rho}'}$ being the unit vector in the direction from the spin $\vec{\rho}$ to $\vec{\rho}'$, see Fig.~\ref{fig:intro}(d,e). The dynamics of the N\'{e}el vector is governed by the equation, obtained within the harmonic expansion of Eq.~\eqref{eq:LL} with respect to $\epsilon$ 
\begin{subequations}\label{eq:eq-motion}
\begin{equation}\label{eq:n-dyn}
\vec{n} \times \left[ \dfrac{M_\textsc{s}^2}{\gamma_0^2\Lambda} \partial_{tt}\vec{n} - A \Delta\vec{n} -K n_z \vec{e}_z + D \nabla \times \vec{n} \right] = 0
\end{equation}
with $\vec{n} \equiv \vec{n}(\vec{r}, t)$ being a continuum (micromagnetic) counterpart of $\vec{n}_{\vec{\rho}}$, $A = \mathscr{J}S^2/(2a_0)$ is the exchange stiffness, $K = \mathscr{K}S^2/(2a_0^3)$ is the anisotropy constant and $D = dS^2/a_0^2$ is the micromagnetic DMI constant. The saturation magnetization of each sublattice is $M_\textsc{s} = g\mu_\textsc{b}S/(2a_0^3)$ with $g$ being Land\'{e} factor and $\mu_\textsc{b}$ is the Bohr magneton and $\gamma_0 = g\mu_\textsc{b}/\hbar$ is the gyromagnetic ratio. The characteristic scales of the magnetic field for $K> 0$ are the spin-flop field $B_\text{sf} =  \sqrt{\Lambda K}/M_\textsc{s}$ with $\Lambda = 6\mathscr{J}S^2/a_0^3$ being the constant of the uniform exchange and spin-flip field $B_\text{x} = \Lambda/M_\textsc{s}$, see Appendix~\ref{app:slasi} and~\ref{app:spinflop} for comparison with spin-lattice simulations. The critical DMI value is $D_c = 4\sqrt{AK}/\pi$. 

The complete formulation of the micromagnetic problem includes boundary conditions for the vector $\vec{n}$. A conventional way to obtain the boundary conditions within the model of multiple sublattices is to consider the difference in torques acting on the boundary spins and bulk spins~\cite{Stamps87,Huang17c,Ghader19a} and matching the equations of motion for both of them. 
The linear in $\epsilon$ analysis of the discrete equations of motion Eq.~\eqref{eq:LL} for $(100)$, $(110)$ and $(111)$ surfaces provides the following boundary conditions
\begin{equation}\label{eq:bc}
\vec{n} \times \left[2  A (\vec{\hat{\nu}} \cdot \nabla) \vec{n} - D \vec{\hat{\nu}}\times\vec{n} \right] = 0 
\end{equation}
with $\vec{\hat{\nu}}$ being the surface normal. We note that when considering other crystallographic cuts, Eq.~\eqref{eq:bc} is not changed within the linear approximation.
\end{subequations}

\begin{figure*}
\includegraphics[width=\linewidth]{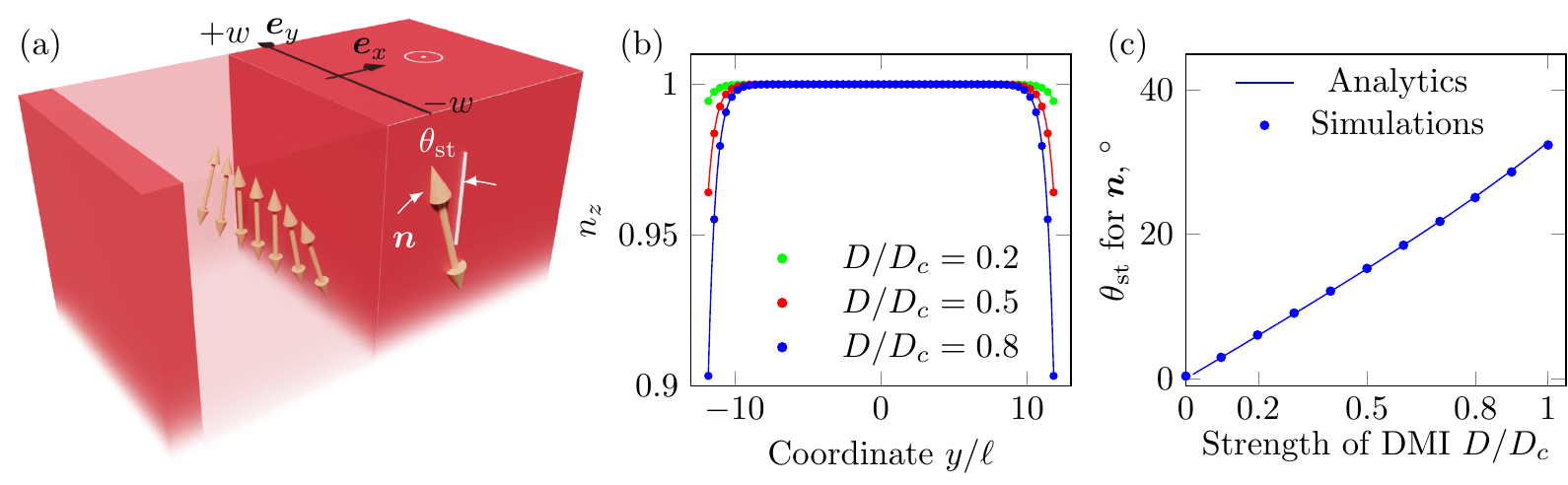}
\caption{\textbf{Ground state of a chiral antiferromagnetic slab.} (a) Schematic representation of the changes of the ground state in confined AFM sample (width $2w<\infty$ along $\vec{e}_y$ direction). At side faces, the N\'{e}el vector is tilted with the surface twist angle $\theta_\text{st}$. The ground state in the interior of the domain is shown within the transparent part of the sample. (b) Vertical component of the N\'{e}el vector at the center of the sample along the y-axis: for different strength of the bulk DMI $D$. Symbols and solid lines correspond to simulations and analytics (Eq.~\eqref{eq:ground-state}), respectively. (c) Surface twist angle $\theta_\text{st}$ for different strength of the bulk DMI. The simulations are carried out for a rectangular slab of $120\times 120\times 160$ spins ($24\ell\times 24\ell\times 32\ell$) with the magnetic length $\ell = 5a_0$.
}
\label{fig:ground}
\end{figure*}

In contrast to antiferromagnets, the influence of boundary conditions is well addressed in ferromagnetism. The behavior of ferromagnets is governed by the Rado--Weertman boundary conditions~\cite{Rado59,Labrune95,Hubert09,Kruglyak14,Busel18} with the DMI related term~\cite{Rohart13}. The lateral confinement alters the shape of chiral domain walls~\cite{Muratov17} and leads to the non-reciprocal domain wall dynamics~\cite{Zhang18c}. The surface twist of the ground state~\cite{Rohart13,Meynell14} and vortex-like textures at the surface~\cite{Rohart13,Wilson13,Meynell14,Luo14,Hals17,Raeliarijaona18} are observed in chiral ferromagnets. This behavior of antiferromagnetic solitons is not known yet and will be addressed in the following.

To study static properties of the ground state and magnetic solitons, we analyze the micromagnetic energy. The continuum micromagnetic functional of the potential energy for the $\sigma$-model, corresponding to the spin-lattice Hamiltonian Eq.~\eqref{eq:hamiltonian} reads
\begin{equation}\label{eq:energy}
\begin{split}
E & = \int \mathscr{E} \mathrm{d}\vec{r},\\
\mathscr{E} & = A (\partial_i \vec{n})(\partial_i \vec{n}) - K n_z^2 + D \vec{n}\cdot \nabla \times \vec{n}.
\end{split}
\end{equation}
As follows from Eq.~\eqref{eq:order-params}, the magnetization $\vec{m}$ is a slave variable,  $\vec{m} = -M_\textsc{s}/(\gamma_0\Lambda) \vec{n}\times \partial_t\vec{n}$. Unlike spin chains~\cite{Ivanov95e,Tveten16}, the symmetry breaking term in~\eqref{eq:energy} is absent. This is a consequence of the possibility to derive the micromagnetic model in a spatially-symmetric way in two-~\cite{Komineas98,Komineas20b} and three dimensions. In crystals with the symmetry lower than the simple cubic one, the symmetry-breaking term appears as the homogeneous DMI~\cite{Dzyaloshinsky57,Dzyaloshinsky58}. The dynamic equation~\eqref{eq:n-dyn} and boundary conditions~\eqref{eq:bc} can be recovered by the variation of the Lagrangian 
\begin{equation}\label{eq:lagranian}
\mathcal{L} =  \frac{M_\textsc{s}^2}{\gamma_0^2\Lambda} \int \left(\partial_t \vec{n} \right)^2 \mathrm{d}\vec{r} - E
\end{equation}
taking into account orthogonality of $\vec{m}$ and $\vec{n}$ within $\mathcal{O}(\epsilon)$.

A similar procedure can be applied to other types of lattices to derive a direct correspondence between the parameters of the micromagnetic model and the spin lattice parameters including the boundary conditions.

\subsection{Ground state}
\label{sec:ground}

The ground state of an \textit{achiral} bipartite antiferromagnet corresponds to the direction of the order parameter along the easy axis of the anisotropy $\vec{e}_z$. This is also true for the boundary spins in the absence of DMI. The presence of the bulk DMI alters the order parameter upon approaching the side faces. The order parameter acquires a tilt at the face surfaces, indicated with a surface twist angle $\theta_\text{st}$ in the schematics in Fig.~\ref{fig:ground}(a). To describe this twist, we consider a slab of the width $2w$ along the $y$ axis assuming the origin in the center of the sample. Using the parametrization $\vec{n} = \{\sin\theta\cos\phi, \sin\theta\sin\phi,\cos\theta \}$ in Cartesian reference frame with $\theta=\theta(y) = \arccos n_z$ and $\phi = \phi(y)$ being polar and azimuthal angles, see Fig.~\ref{fig:intro}(e), the minimum of the energy~\eqref{eq:energy} is reached with 

\begin{equation}\label{eq:ground-state}
\begin{split}
n_z & = \tanh \left[\arctanh\sqrt{1 - \left(\frac{2}{\pi}\frac{D}{D_c}\right)^2} + \frac{w - |y|}{\ell} \right],\\
\phi & = -\frac{\pi}{2},
\end{split}
\end{equation}
where the requirement $\partial_y \theta(|y|=w) = D/(2A)$ comes from the boundary conditions~\eqref{eq:bc}. We note that for the case of the AFM order parameter (director), the states with $(\theta,\phi)$ and $(\pi - \theta, \phi \pm \pi)$ are equivalent. The ground state is significantly altered by the boundary at the distance about $2\ell$. The order  parameter possesses a twist by the angle $\theta_\text{st} = \arccos n_z(w)$ at the boundary, determined by the strength of the bulk DMI, see lines in Fig.~\ref{fig:ground}(b). The surface twist angle $\theta_\text{st}$ grows almost linearly with $D$ and can reach 30$^\circ$ when approaching the critical DMI $D_c$, see line in Fig.~\ref{fig:ground}(c). The surface twist angle is determined by the relation $D/D_c$ only because the critical DMI $D_c$ holds the exchange and anisotropy scales in the first term in~\eqref{eq:bc}, while $D$ governs the twist itself. The observed effect is similar to the boundary twists observed in ferromagnets~\cite{Rohart13,Meynell14}. 

The analytical results shown in Fig.~\ref{fig:ground}(b,c) are confirmed by spin-lattice simulations, performed for a rectangular slab containing $120\times 120\times 160$ spins with the magnetic length $\ell = 5a_0$, see Appendix~\ref{app:slasi} for details. 

\subsection{Domain wall}
\label{sec:dw}

\begin{figure*}[t]
\includegraphics[width=\linewidth]{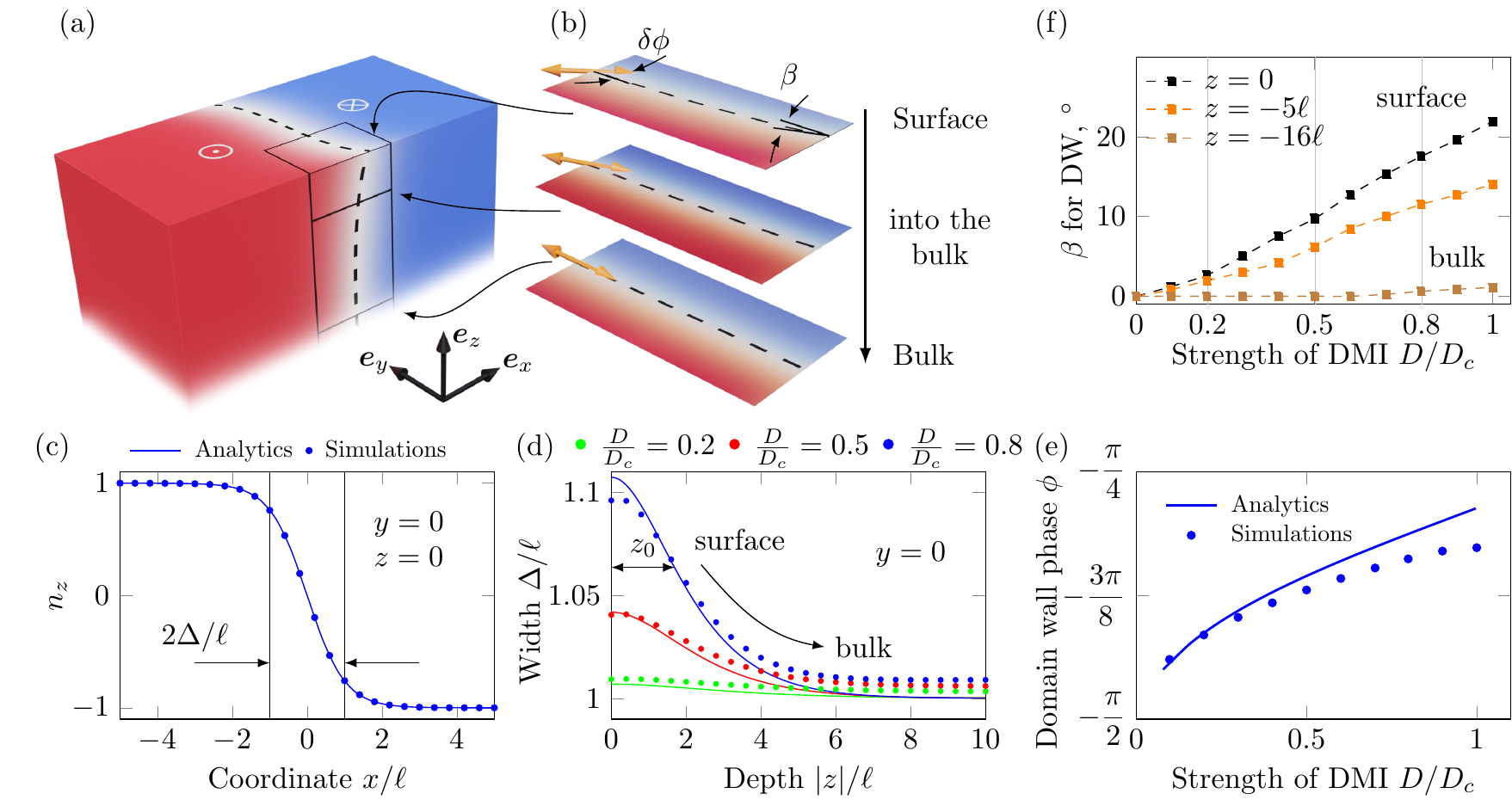}
\caption{\textbf{Chiral domain walls in antiferromagnetic slabs.} {(a)} Schematics of the domain wall (DW) in a confined sample of a constant width $2w_y < \infty$ along $\vec{e}_y$ axis. {(b)} At the sample's edge, the domain wall acquires a tilt indicated with the angle $\beta \neq 0$. The tilt vanishes ($\beta = 0$) in the bulk. The panel (b) shows the simulation results for $D/D_c = 0.8$ with geometry and other material parameters same as in Fig.~\ref{fig:ground}. Dashed line represents the center of the domain wall. Arrows schematically show the surface twist $\delta\phi$ of the domain wall phase $\phi$ far from the side faces. (c) Domain wall profile with $p = -1$ is shown for the component of the order parameter $n_z$ along $\vec{e}_x$ axis. {(d)}  Domain wall width $\Delta$ expands at the surface $z = 0$ in comparison with the bulk. Solid lines and symbols correspond to the analytics~\eqref{eq:dw-width-equation} and simulations, respectively.  (e) Domain wall phase $\phi = -\pi/2-\delta\phi$ [see panel (b)] at the sample's surface far from edges. {(f)} Tilt angle $\beta$ for different DMI. Dashed lines are guides to the eye.
}
\label{fig:dw-tilt}
\end{figure*}

A G-type antiferromagnet supports translational or so-called phase domain states~\cite{Cheong20}, where the wall separates domains with the swapped order of sublattices on the atomistic level [Fig.~\ref{fig:intro}(b)]. These domains schematically colored in red and blue, are shown in Fig.~\ref{fig:dw-tilt}(a). We consider the translational domain wall initially located in $yz$ plane with the origin of the reference frame lying at the center of the top surface. In the absence of DMI, the domain wall plane is flat and perpendicular to the side faces of the sample being quasi-1D texture with $\cos\theta = p \tanh (x/\Delta)$ and $\phi = \text{const}$, where $\Delta = \ell$ is the domain wall width and $p = \pm 1$ is the domain wall polarity. In the absence of additional anisotropies or chiral interactions, the phase $\phi$ is not determined. A finite $D$ leads to the preferred domain wall chirality. Namely, the last term in the energy density $\mathscr{E} = A(\partial_x \theta)^2 + K\sin^2\theta + D \sin\phi \partial_x \theta$ forces the stabilization of a Bloch-type domain wall with the favorable chirality $\sin\phi = C = \sign (pD)$.

A lateral confinement of the domain wall leads to its deformation (bent and broadening) and change of its internal structure, see Fig.~\ref{fig:dw-tilt}(a,b). We start with the description of the internal structure of the domain wall near the top surface and far from the side faces of the sample. The domain wall profile can be described using a 2D Ansatz $\cos\theta(x,z) = p\tanh [x/\Delta(z)]$ and $\phi(z) = C\pi/2 + \delta\phi(z)$ with $x \in (-\infty, \infty)$ and $z \in (-\infty, 0]$, see Fig.~\ref{fig:dw-tilt}(c). The twist of the domain wall near the top surface reads
\begin{equation}\label{eq:dw-twist}
\delta\phi(z) = \phi_0 \exp \frac{z}{\lambda}
\end{equation}
with the parameter $\lambda$ characterizing the penetration depth. The boundary conditions~\eqref{eq:bc} require $\partial_z \phi(0) = D/(2A)$ and $\partial_z\Delta(0) = 0$ at the top surface, while free boundary conditions at $z = -\infty$ are assumed for both functions. The substitution of the Ansatz in~\eqref{eq:energy} gives the following effective energy density after integration along $x$ axis:
\begin{equation}\label{eq:dw-energy-density}
\begin{split}
\mathscr{E}_\text{dw} &= 2K\Delta + A \frac{12+\pi^2(\partial_z\Delta)^2}{6\Delta} + \pi p CD \cos \left( \frac{\lambda D}{2A} e^{z/\lambda} \right)\\
& - \frac{D^2}{2A}\Delta e^{z/\lambda} (2-e^{z/\lambda}),\quad z\le 0.
\end{split}
\end{equation}
Here, the first and the second terms correspond to the anisotropy and the part of the exchange energy density dependent on the derivatives of $\theta$. The third term represents the part of the DMI energy density related to the $\partial_x\theta$. It determines $C$ in the same way as for the Bloch domain wall in the infinite medium. The last term in~\eqref{eq:dw-energy-density} originates from parts of the exchange and DMI energy densities related to the $\partial_z\phi$. The domain wall width $\Delta$ is determined by the following variational equation, independent of $C$ and $p$:
\begin{equation}\label{eq:dw-width-equation}
\begin{split}
\frac{\pi^2}{6} \frac{A}{\Delta^2} &  \left[ 2\Delta \partial_{zz}\Delta - (\partial_z \Delta)^2 \right]  + \frac{2A}{\Delta^2} \\
&+ \frac{D^2}{2A}e^{z/\lambda}(2-e^{z/\lambda}) = 2K,\quad z\le 0.
\end{split}
\end{equation}
with $\lambda$ being an unknown parameter to be found from the energy minimization, see Appendix~\ref{app:dw} for details. We find that the domain wall becomes wider near the top surface up to about 10\% in comparison with the bulk value, see lines in Fig.~\ref{fig:dw-tilt}(d). The function $\Delta$ has a Gaussian-like shape. Its characteristic half-width $z_0$ of the $\Delta$ is $1.7\ell$ for $D_c$ and grows with the decrease of $D$ being $2.7\ell$ at $D = 0.2D_c$. The twist of the domain wall at the top surface, $\phi$, increases with the the strength of the DMI, see line in Fig.~\ref{fig:dw-tilt}(e). The penetration depth of the phase $\lambda$ behaves in a similar way as $z_0$. It equals $\ell$ for $D=D_c$ and becomes $2.1\ell$ for $D = 0.2D_c$. Note, that the increase of $\lambda$ and $z_0$ for smaller values of DMI are accompanied by a rapid reduction of $\phi_0$ and width $\Delta(0)$ at the top surface. 

We elaborate the model of the AFM domain wall in a slab using spin-lattice simulations. They show a reasonable quantitative agreement with the analytical predictions, see symbols in Fig.~\ref{fig:dw-tilt}(c,d,e). The value of $\Delta$ for $|z| \gtrsim 6\ell$ obtained in simulations is slightly larger than $\ell$ due to effects of discreteness and is reduced with smaller $\epsilon$ used for numerical investigations. In addition, we numerically analyze the domain wall behavior near the sample edges. The domain wall plane possesses a twist, which is observed as an ``S-shaped'' profile at the top surface, see Fig.~\ref{fig:dw-tilt}(a,b) for schematics and simulations. We characterize this distortion by the angle $\beta$ with respect to the edge normal within the plane of the top surface. While $\beta = 0$ in the absence of DMI, a finite $D$ leads to the increase of $\beta$ up to $20^\circ$, see Fig.~\ref{fig:dw-tilt}(f). There is a slow reduction of $\beta$ to the equilibrium value $\beta_\text{eq} = 0$ far from the sample's edges with about of 60\% of the surface value at the depth $z = -5\ell$. Thus, taking into account the variation of the domain wall width near the surface and bend at the edges of the sample, the bulk-like properties of this texture are preserved for samples significantly thicker and wider than $10\ell$.

\subsection{Skyrmion}
\label{sec:skyrmion}

\begin{figure*}[t]
\includegraphics[width=\linewidth]{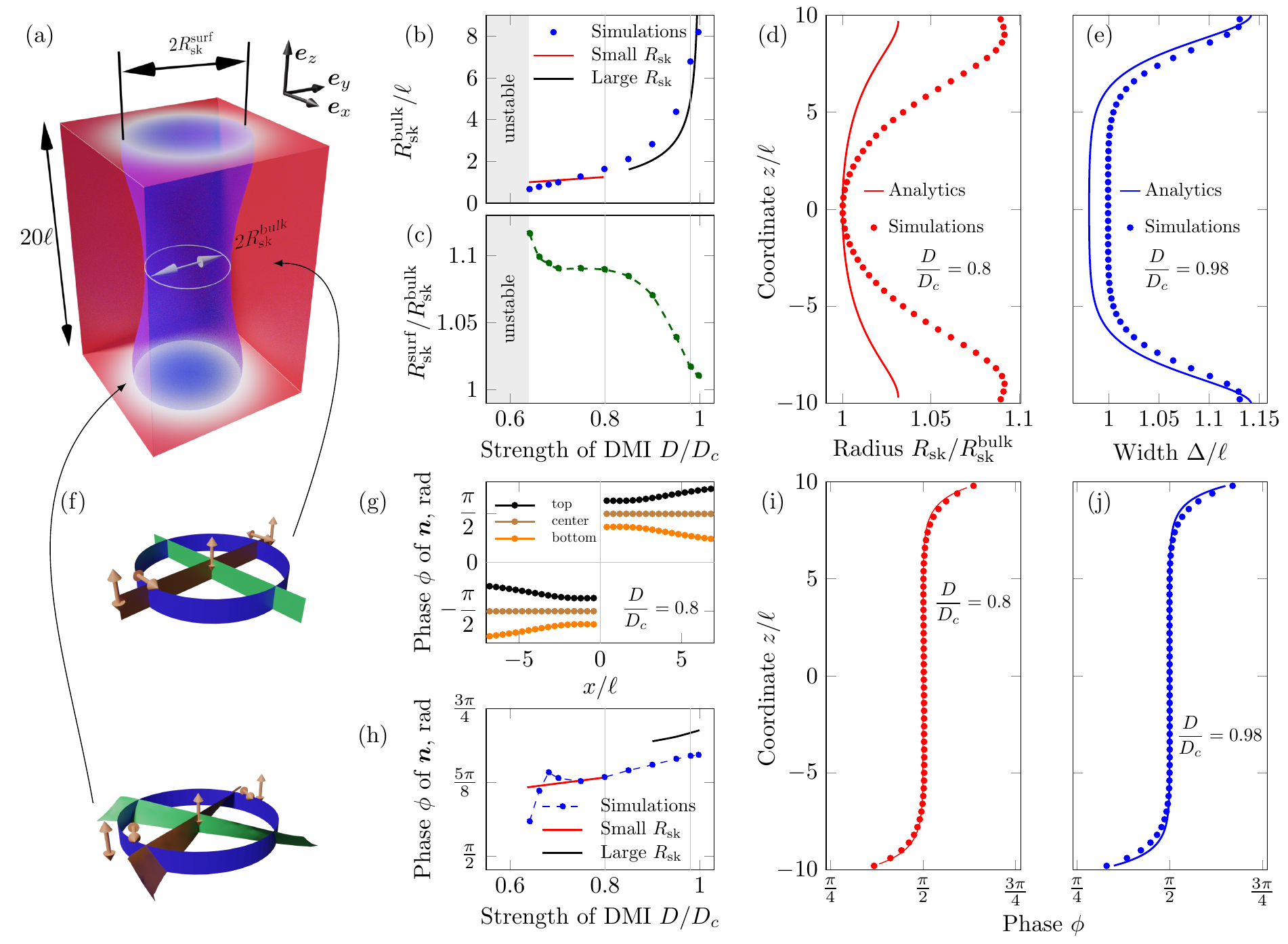}
\caption{\textbf{Skyrmion in antiferromagnetic slabs.} {(a)} Schematics of a skyrmion in a thick AFM slab. Its radii at the surface and in the bulk are marked as $R_\text{sk}^\text{surf}$ and $R_\text{sk}^\text{bulk}$, respectively. In the simulations we consider a slab consisting of $150\times 150\times 100$ spins with $\ell = 5a_0$. %
{(b,c)} Skyrmion radius as a function of the DMI strength. ``Unstable'' marker shows the region where the skyrmion of small radius is unstable. Symbols and solid lines correspond to simulations and analytics~\eqref{eq:small-sk-radius-bulk},~\eqref{eq:large-sk-radius-bulk}, respectively. Dashed line is guide to the eye. 
{(d)} Skyrmion radius for different axial cross-sections of the sample. Solid lines correspond to the Ansatz of the skyrmion of small radius. 
{(e)} Width of the skyrmion of large radius measured at $R_\text{sk}^\text{bulk}$. 
{(f)} Skyrmion isosurfaces $n_{x,y,z}(\vec{r}) = 0$ in the center and at the bottom of the slab. While this is a purely Bloch skyrmion in the bulk, the $\vec{n}$ is tilted when approaching the surface. 
{(g,h)} Phase $\phi$ at the top surface for different strengths of the DMI at the distance $x = R_\text{sk}^\text{surf}$ from the origin. Symbols and solid lines correspond to simulations and analytics, respectively. Dashed line is guide to the eye.
(i,j) Phase $\phi$ of $\vec{n}$ at the distance $R_\text{sk}^\text{bulk}$ from the origin for skyrmions of small and large radius, respectively. Notations are the same as in panel {(d)}.
}
\label{fig:skyrmion}
\end{figure*}

A skyrmion is a chiral texture stabilized by the DMI~\cite{Ivanov95e,Bogdanov02a}. In this section, we consider impact of the 3D confinement on individual skyrmions in a chiral AFM slab with the focus on the modification of the shape and phase of the skyrmion, see Fig.~\ref{fig:skyrmion}(a). Even for ferromagnets, a rigorous description of skyrmions of small~\cite{Komineas20a} and large~\cite{Komineas19} radius is a complicated task, which is usually addressed by asymptotic analysis or numerically. Often, models of circular domain walls or numerical integration are utilized, which allows to explain current-driven dynamics~\cite{Velkov16,Shen18b,Komineas20b} and excitations~\cite{Kravchuk19a}. To address a 3D skyrmion texture, we describe them qualitatively using an axially symmetric Ansatz $\theta = \theta(r,z)$ and $\phi = \phi(\chi,z)$. To highlight the peculiarities of the confined geometry, we consider a semi-infinite slab with $z \in (-\infty, 0]$ in analytics and a sufficiently thick and wide rectangular box in simulations.

We start with the analysis of skyrmions of \textit{small radius}. Their bulk properties can be addressed with the linear Ansatz~\cite{Bogdanov94}
\begin{equation}\label{eq:linear-sk}
\begin{split}
\theta_\text{sk\,sm}(r) & = \begin{cases}
\pi \left(1 - \dfrac{r}{2R_\text{sk\,sm}^\text{bulk}}\right) & r \le 2R_\text{sk\,sm}^\text{bulk},\\
0 & r > 2R_\text{sk\,sm}^\text{bulk},
\end{cases}\\
\phi(\chi) & = \chi + \tilde{\phi}
\end{split}
\end{equation}
where the phase $\tilde{\phi} = \text{const}$. Here and below we use the definition of the skyrmion radius $R_\text{sk}$ as $\theta(R_\text{sk}) = \pi/2$. The energy~\eqref{eq:energy} integrated with~\eqref{eq:linear-sk} reaches minimum at
\begin{equation}\label{eq:small-sk-radius-bulk}
R_\text{sk\,sm}^\text{bulk} = \ell\frac{\pi}{2} \frac{|D|}{D_c},\quad \sin\tilde{\phi} = C = \sign D,
\end{equation}
see Appendix~\ref{app:sk} for details. This corresponds to a Bloch skyrmion with the radius linearly growing with $D$, see red line in Fig.~\ref{fig:skyrmion}(b). The skyrmion of small radius in the sample with a sufficiently large lateral size is influenced only by the top surface $z = 0$. To address this spatial confinement in the vertical dimension, we modify the Ansatz~\eqref{eq:linear-sk} adding the dependence on the longitudinal coordinate $z$, namely $R_\text{sk\,sm}^\text{bulk} \to R_\text{sk\,sm}(z)$ and using the phase $\phi(\chi,z) = \chi + C\pi/2 + \delta\phi(z)$ with the definition of $\delta\phi(z)$ according to~\eqref{eq:dw-twist}. The boundary conditions~\eqref{eq:bc} lead to $\partial_z R_\text{sk\,sm}(0) = 0$ and $\partial_z\phi(0) = D/(2A)$. Substitution of this Ansatz into energy~\eqref{eq:energy} and integration along the radial direction allows to obtain a variational equation for $ R_\text{sk\,sm}(z) $ with the parameter $\lambda$ similar to the case of the domain wall~\eqref{eq:dw-width-equation}, see Appendix~\ref{app:sk} for details. The solution of the obtained equation shows that the skyrmion possesses a Gaussian-like bottle-neck shape. The skyrmion is narrow in the bulk and becomes wider when approaching the surface, see red line in Fig.~\ref{fig:skyrmion}(d). The skyrmion shows a mixed Bloch--N\'{e}el texture at the surface due to the twist governed by the boundary conditions, see red line in Fig.~\ref{fig:skyrmion}(i). Note, that a similar shape distortion is observed for vortices in easy-plane ferromagnets with surface anisotropy~\cite{Pylypovskyi14, Pylypovskyi15a}.

\emph{Large radius} skyrmions in bulk samples can be described as circular domain walls using the Ansatz~\cite{Kravchuk19a}
\begin{equation}\label{eq:circular-sk}
\cos\theta_\text{sk\,lar} = \tanh \frac{r - R_\text{sk\,lar}^\text{bulk}}{\Delta},\quad \phi(\chi) = \chi + \tilde{\phi}
\end{equation}
assuming $\Delta \ll R_\text{sk\,lar}^\text{bulk}$. The energy~\eqref{eq:energy} reaches minimum with this Ansatz at~\footnote{Note, that Ansatz~\eqref{eq:large-sk-radius-bulk} also works for small radius skyrmions~\cite{Kravchuk19a}.}
\begin{equation}\label{eq:large-sk-radius-bulk}
\begin{split}
R_\text{sk\,lar}^\text{bulk} & \approx \frac{|D|/D_c}{\sqrt{1-D^2/D_c^2}},\quad \Delta \approx \dfrac{|D|}{D_c},\\
\sin\tilde{\phi} & = C = \sign D,
\end{split}
\end{equation}
see black line in Fig.~\ref{fig:skyrmion}(b). While it is expected that the skyrmion radius should not be significantly influenced by the sample's boundary, its width $\Delta$ and phase $\phi$ are altered due to confinement. Therefore, the 3D texture can be described by the replacement $\Delta \to \Delta(z)$ and $\phi(\chi) \to \phi(\chi,z) = \chi + C\pi/2 + \delta\phi(z)$ in~\eqref{eq:circular-sk}. We find that the structure of the circular domain wall stabilized by the DMI is similar to the straight one, considered in Sec.~\ref{sec:dw}, see Fig.~\ref{fig:skyrmion}(e,j).

To obtain a quantitative description of the skyrmion shape in a confined geometry, we elaborate the above analytics by spin-lattice simulations performed for a slab consisting of $150\times 150\times 100$ spins with $\ell = 5a_0$. The stability of the skyrmion is influenced by the discreteness of the system and 3D shape of the texture. We find that the skyrmion can be relaxed for $D \gtrsim 0.64D_c$ with the smallest bulk radius $R_\text{sk}^\text{bulk} \approx 3.4a_0$, see Fig.~\ref{fig:skyrmion}(b). The skyrmion radius in the bulk grows with $D$ up to $8.2\ell$ at $0.997D_c$. The largest size of the skyrmion is limited by the lateral dimensions of the sample~\cite{Rohart13}. The ratio of the skyrmion radius at the surface and in the bulk $R_\text{sk}^\text{surf}/R_\text{sk}^\text{bulk}$ found numerically is in agreement with the analytical model: it is large for small $D$ and reduces to $1$ when approaching $D_c$, see Fig.~\ref{fig:skyrmion}(c). The skyrmion possesses a complex structure in the bulk as well as at the surface. The longitudinal profile shows two maximal radii at the distance of about $\ell$ from the top and bottom surfaces, see symbols in Fig.~\ref{fig:skyrmion}(g). The phase of $\vec{n}$ possesses a radially dependent asymmetric surface twist, which is changed with DMI, see Fig.~\ref{fig:skyrmion}(g,h). While analytics quantitatively capture the spatial profile of the phase $\phi$ (Fig.~\ref{fig:skyrmion}(i,j)), only qualitative agreement is obtained for the bulk and surface skyrmion radii.

\section{Conclusions}
\label{sec:conclusions}

We derive a nonlinear $\sigma$-model with boundary conditions for a uniaxial chiral antiferromagnet of G-type with a simple cubic lattice and DMI of surface and bulk types. We establish a correspondence between the spin lattice and micromagnetic parameters relying on the approach with the N\'{e}el vector order parameter $\vec{n}$. The transition between spin lattice and micromagnetic models requires six auxiliary fields, determined by the spatial derivatives of the N\'{e}el vector. 
The micromagnetic boundary conditions for the N\'{e}el vector match the variational derivation from the micromagnetic Lagrangian and are similar to the Rado--Weertman ones with the DMI term for ferromagnets~\cite{Rado59,Hubert09,Rohart13}. The difference lies in the symmetry of the order parameter: the states of vector--director $\vec{n}$ and $-\vec{n}$ are indistinguishable. A procedure described here for the case 3D antiferromagnets with a simple cubic lattice can be straightforwardly extended to other types of lattices. 

The obtained model is applied to analyze the ground state and magnetic solitons in a spatially confined sample. In this discussion, we focused on the case when the AFM slabs possesses a bulk DMI. The order parameter in the ground state acquires a chiral surface twist at the boundaries due to the lack of neighboring spins and competing exchange and DMI energy terms. Depending on the DMI strength, the value of the surface twist angle can reach up to 30$^\circ$. The noncollinear textures, such as domain walls and skyrmions, become modified near the boundary with the characteristic penetration depth of about 5 magnetic lengths. The domain wall being laterally constrained, possessed an S-shaped bend at the surface. Both, the domain wall and skyrmion become of the mixed, Bloch--N\'{e}el type at the surface. The DMI forces the skyrmions and domain walls to become broader near the surface. In particular, for skyrmions of small radius, the radius becomes 10\% larger when approaching the face of the sample. 

The here discussed impact of the confined geometry and DMI on the static magnetic textures provides an estimate for the minimal dimensions of AFM samples hosting chiral magnetic solitons with bulk-like properties. Furthermore, we note that the change of the size of the textures when approaching the boundaries is expected to alter their dynamic properties. In this respect, the presented model can be applied for perspective design of AFM racetracks and description of AFM textures in structured samples~\cite{Hedrich20}.

\section*{Acknowledgments}
Authors thank Prof. Patrick Maletinsky, Natascha Hedrich, Dr. Kai Wagner and Dr. Brendan J. Shields (University of Basel) for fruitful discussions. This work was financed in part via the German Research Foundation (DFG) grants MA 5144/22-1, MC 9/22-1, MA 5144/24-1, Alexander von Humboldt Foundation (Research Group Linkage Programme), and by the Ministry of Education and Science of Ukraine (Project 19BF052-01).
 
\appendix

\section{Description of the spin lattice}
\label{app:spinlattice}

To describe a G-type antiferromagnet, we split the lattice into groups of octamers. Within a single octamer enumerated by the vector index $\vec{\rho} = \{i,j,k\}$, the spins are labeled by the Latin letters $\vec{A}_{\vec{\rho}}$,\ldots,$\vec{H}_{\vec{\rho}}$, see Fig.~\ref{fig:intro}(a). Then, the coordinate of each spin is $\vec{\rho} + \{ \alpha, \beta, \gamma \}$, where $\alpha$, $\beta$ and $\gamma$ running  0,~1. In the following, we use $\eta,\zeta \in \{x,y,z\}$ for subscripts and spatial derivatives and $\eta,\zeta \in \{\alpha,\beta,\gamma\}$ in exponents. For example, $\sum_{\eta} (-1)^\eta \vec{u}_\eta = (-1)^\alpha \vec{u}_x + (-1)^\beta \vec{u}_y + (-1)^\gamma \vec{u}_z$ takes eight different values for different $\alpha,\beta,\gamma$. The single index $\vec{\rho}$ is omitted for simplicity. Then, the unit magnetic moment within an octamer reads
\begin{equation} \label{eq:octamer}
\begin{aligned}
\vec{\mu}_{\vec{\rho} + \{\alpha,\beta,\gamma\}} & = \vec{m} + (-1)^\xi \vec{n} + \vec{p}(\alpha,\beta,\gamma),\\
\vec{p}(\alpha,\beta,\gamma) & = (-1)^{\xi + \eta} \vec{u}_\eta + (-1)^\eta \overline{\vec{u}}_\eta
\end{aligned}
\end{equation}
where $\xi = \alpha+\beta+\gamma$, the triple $\{\alpha,\beta,\gamma\}$ enumerate the spin within the given octamer and Einstein summation rule is used. In the following, we apply a multiscale analysis to  describe the micromagnetic transition from the spin lattice approach~\eqref{eq:model-g} using $\epsilon = \sqrt{\mathscr{|K|}/\mathscr{J}}$  as a scaling parameter. 

To describe the behavior of the spin system in the continuum limit, the following relations for the neighboring spins along $\eta$ direction are used:
\begin{equation}\label{eq:spin-expansion}
\begin{aligned}
\vec{V}_{\vec{\rho} \pm \vec{\Delta}_\eta} & = \vec{V}(\vec{r}) \pm 2\epsilon \ell \partial_\eta \vec{V}(\vec{r}) + 2\epsilon^2\ell^2 \partial_\eta ^2 \vec{V}(\vec{r}),\\
\vec{V} & = \vec{A},\ldots,\vec{H}.
\end{aligned}
\end{equation} 
Considering slow spatial and temporal variations of the magnetic moments, we rewrite the equations of motion~\eqref{eq:LL} using dimensionless time $\tau = \epsilon \Omega t$ with $\Omega = \mathscr{J}S/\hbar$. 

The rescaled anisotropy and DMI coefficients are $k_0 = \epsilon^2$ and $\delta_0 = \epsilon d_0$. This also implies that $\vec{m}$ and auxiliary fields $\vec{u}_\eta$ and $\overline{\vec{u}}_\eta$ are of the order of $\epsilon$ for the N\'{e}el ground state in the bulk.

The linear expansion of~\eqref{eq:LL} at the site $\{\alpha,\beta,\gamma\}$ reads
\begin{equation}\label{eq:eqmot-linear}
\begin{aligned}
\dot{\vec{n}} =&  - 12\vec{m}\times \vec{n} + 2\ell\vec{n}\times (-1)^\eta \partial_\eta \vec{n}\\
& + 4\vec{n} \times \left[ \vec{p} + (-1)^\eta \overline{\vec{u}}_\eta \right],
\end{aligned}
\end{equation}
where overdot means the derivative with respect to $\tau$. The expression~\eqref{eq:eqmot-linear} represents eight equations with respect to different values of $\alpha,\beta,\gamma$. The solution of \eqref{eq:eqmot-linear} within each octamer is given by Eq.~\eqref{eq:order-params}. It provides the relations between the primary and auxiliary vector fields, describing each octamer.

The harmonic expansion of Eq.~\eqref{eq:LL} provides equations of motion for $\vec{m}$. For the given $\{\alpha,\beta,\gamma\}$, they read
\begin{widetext}
\begin{equation}\label{eq:eqmot-harmonic}
\begin{aligned}
\dot{\vec{m}} & + (-1)^{\xi+\eta}\dot{\vec{u}}_\eta + (-1)^\eta \dot{\overline{\vec{u}}}_\eta  = \underbrace{- \vec{n} \times (n_z \vec{e}_z)}_\text{anisotropy} \underbrace{+ 2\ell d_0 \vec{n}\times \left[ \nabla \times \vec{n} \right]}_\text{DMI} \underbrace{ -\ell^2 \vec{n}\times \Delta \vec{n} - 2 \vec{m} \times \vec{p} }_\text{exchange} \\
 & \underbrace{- 2 \left[ \vec{m} + \vec{p} \right] \times \left[ \vec{p} + (-1)^{\xi+\eta}\vec{u}_\eta\right]  - (-1)^{\xi+\zeta} \ell \vec{n} \times \partial_\zeta \left[ \vec{m}+ (-1)^{1+\xi+\eta+\delta[\eta,\zeta]}\vec{u}_\eta + (-1)^{\eta+\delta[\eta,\zeta]}\overline{\vec{u}}_\eta \right]}_\text{exchange},
\end{aligned}
\end{equation}
\end{widetext}
where $\delta[\eta,\zeta]$ is the Kronecker delta with respect to symbols $\eta$ and $\zeta$. Summation of~\eqref{eq:eqmot-harmonic} for all possible values of $\alpha,\beta,\gamma$ within each octamer and excluding $\vec{m}$ leads to Eq.~\eqref{eq:n-dyn}. 

The boundary conditions can be rigorously obtained from the equations of motion of the boundary spins. The equations of motion within the continuum limit are the same in the bulk and at the surfaces, while the spins have different number of neighbors and experience different torques. The boundary conditions arise as the match between boundary and surface torques. For example, considering a $(111)$ surface with the normal vector $\vec{\hat{\nu}} = \{1,1,1\}/\sqrt{3}$, the boundary spin is $\vec{H}_{\vec{\rho}}$ with absent neighbors $\vec{G}_{\vec{\rho} + \{1,0,0\}}$, $\vec{E}_{\vec{\rho} + \{0,1,0\}}$ and $\vec{C}_{\vec{\rho} + \{0,0,1\}}$. This implies
\begin{equation}\label{eq:bc-111}
\begin{split}
\vec{H} \times \left(\vec{G}_{i+1} + \vec{E}_{j+1} + \vec{C}_{k+1}\right)\\ +  d_0 \left( \vec{G}_{i+1}\times {\vec{e}_x} + \vec{E}_{j+1}\times {\vec{e}_y} + \vec{G}_{k+1} \times {\vec{e}_z} \right) = 0.
\end{split}
\end{equation}
Substitution of the expressions for spins~\eqref{eq:octamer} allows to reduce~\eqref{eq:bc-111} to~\eqref{eq:bc}.

The total energy of the $\sigma$-model reads
\begin{equation}\label{eq:energy-tot}
E_\text{tot} = \bigintsss \left( \dfrac{M_\textsc{s}^2}{\gamma_0^2}\dot{\vec{n}}^2 + \mathscr{E} \right) \mathrm{d}\vec{r},
\end{equation}
where $\mathscr{E}$ is the potential energy density introduced in~\eqref{eq:energy}.

\section{DMI of the surface type}
\label{app:surface}

The DMI of the surface type can be obtained for the DMI vector $\vec{d}_{\vec{\rho},\vec{\rho}'} = d \vec{e}_z \times \vec{e}_{\vec{\rho},\vec{\rho}'}$. In this case, the energy of the surface DMI reads
\begin{equation}\label{eq:e-surf-dmi}
E_\textsc{dm}^\text{surf} = h D \int \left[ n_z (\nabla_{xy} \cdot \vec{n}) - (\vec{n}\cdot \nabla_{xy})n_z\right] \mathrm{d}S,
\end{equation}
where $h$ is the sample thickness, $D$ has the same value as for the DMI of the bulk type and the magnetic texture is assumed to be homogeneous along $\vec{e}_z$. The derivation of~\eqref{eq:e-surf-dmi} implies $d = 0$ for $ \vec{e}_{\vec{\rho},\vec{\rho}'} \| \vec{e}_z$. 
This allows to derive the equation of motion and boundary conditions for the N\'{e}el vector similarly to the case of the bulk DMI:
\begin{subequations}
\begin{equation}
\begin{aligned}
\vec{n} \times \Biggl\lbrace \dfrac{M_\textsc{s}^2}{\gamma_0^2\Lambda} \partial_{tt}\vec{n} & - A \Delta\vec{n} -K n_z \vec{e}_z \\
& - D \left[ (\nabla\cdot \vec{n})\vec{e}_z - \nabla n_z \right] \Biggr\rbrace = 0
\end{aligned}
\end{equation}
\begin{equation}
\vec{n} \times \left\{ 2A(\vec{\hat{\nu}}\cdot \nabla) \vec{n} + D\left[ n_z\vec{\hat{\nu}} - (\vec{\hat{\nu}}\cdot \vec{n})\vec{\hat{\nu}} \right] \right\}
\end{equation}
\end{subequations}
with all derivatives within the $\vec{e}_{x,y}$ plane.

\section{Spin-lattice simulations}
\label{app:slasi}

We numerically solve the Landau--Lifshitz equation~\eqref{eq:LL} with the Gilbert relaxation torque $\vec{T}_\text{relax} = \alpha_\textsc{g} \vec{\mu_{\vec{\rho}}} \times \partial_t \vec{\mu}_{\vec{\rho}}$ and $\alpha_\textsc{g}$ being the relaxation constant for the Hamiltonian~\eqref{eq:hamiltonian} using the spin lattice simulator SLaSi~\cite{SLaSi}. To analyze the spin-flop and spin-flip behavior, an additional term $\mathscr{H}_\text{zee} = - \sum_{\vec{\mu}} g\mu_\textsc{b} S \vec{\mu}_{\vec{\rho}} \cdot \vec{B}_\text{zee}$ with $\vec{B}_\text{zee}$ being the external magnetic field is included to~\eqref{eq:hamiltonian}. To model an infinite medium, periodic boundary conditions are applied. We use the following parameters: the spin length $S = 1$, the exchange integral $\mathscr{J} = 2.34\times 10^{-22}$\,J, the constant of a single-ion anisotropy $\mathscr{K} = 9.36\times 10^{-23}$\,J, the Gilbert constant $\alpha_\textsc{g} = 0.5$ to accelerate the relaxation by overdamping, and the absolute value of the DMI vector $d$ is varying from 0 to $2.98\times 10^{-22}$\,J. The integration is performed using the midpoint algorithm at GPU with the time step $\delta t = 0.01$\,ps. The relaxation is performed during 2\,ns. Simulations were carried out using the high performance clusters at the HZDR~\cite{hypnos} and TSNUK~\cite{unicc}.

Figs.~\ref{fig:intro}(b,c) are built taking into account that the geometrical width of the sample is $2w = 24\ell$ and the position of the N\'{e}el vectors at the boundary in simulations correspond to the effective width $2w_\text{eff} = 23.6\ell$. 

\section{External magnetic field}
\label{app:spinflop}
\begin{figure}
\centering
\includegraphics[width=\linewidth]{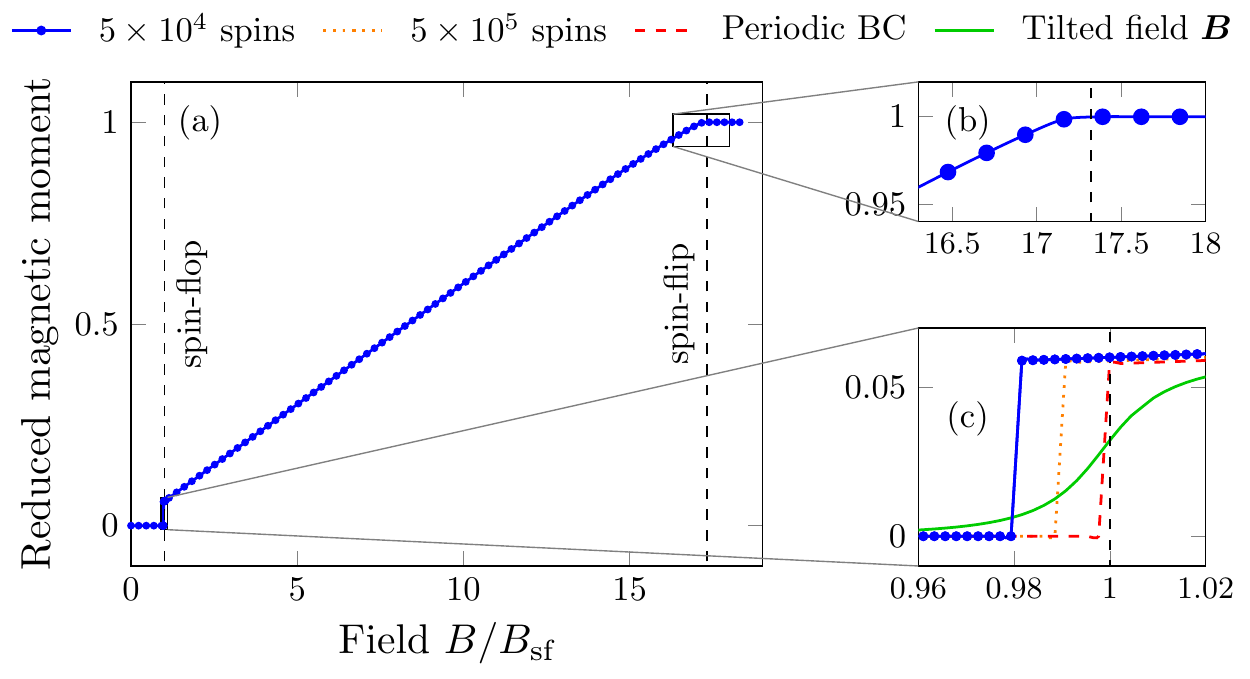}
\caption{\textbf{Spin-flop and spin-flip transitions.} (a) Symbols correspond to spin-lattice simulations and line is guide to the eye. Dashed lines correspond to the analytically found values of the spin-flop and spin-flip fields $B_\text{sf}$ and $B_\text{x}$. Insets (b,c) show zoom of spin-flip and spin-flop regions in (a). Solid blue and dotted orange lines correspond to the samples with dimensions $50\times 50\times 20$ and $100\times 100\times 50$ spins, respectively. Dashed red and solid green lines correspond to simulations with periodic boundary conditions (BC). The field is tilted by $1^\circ$ angle from $\vec{e}_z$ in the $xz$ plane.}
\label{fig:spin-flop}
\end{figure}

The Hamiltonian describing the interaction of the spin lattice with the external magnetic field $\vec{B}_\text{zee}$ reads
\begin{equation}\label{eq:hamiltoninan-zeeman}
\mathscr{H}_\text{zee} = - g\mu_\textsc{b} S \sum_{\vec{\mu}} \vec{\mu}_{\vec{\rho}} \cdot \vec{B}_\text{zee}
\end{equation}
with the corresponding continuum counterpart
\begin{equation}\label{eq:energy-zeeman}
E_\text{zee} = -2 M_\textsc{s} \int \vec{m}\cdot \vec{B}_\text{zee} \mathrm{d}\vec{r}.
\end{equation}
We relaxed the spin lattice exposed to an external magnetic field using two staggered initial states: along and perpendicularly to the anisotropy axis. The energies of the stable states are compared to determine the phase transition. The spin-flop and spin-flip transitions, for the case when $\vec{B}_\text{zee}$ is applied along the anisotropy axis $\vec{e}_z$, are shown in Fig.~\ref{fig:spin-flop}. Fig.~\ref{fig:spin-flop}(c) shows the dependency of the spin-flop field $B_\text{sf}$ on the boundary conditions in simulations. Smaller samples have smaller $B_\text{sf}$, while the sample with periodic boundary conditions, equivalent to the infinite system, shows the exact agreement with theory. We note that the auxiliary fields $\vec{u}_{x,y,z}$ and $\overline{\vec{u}}_{x,y,z}$ do not influence the spin-flop and spin-flip even for the case of finite $\epsilon$ for the homogeneous texture.

\section{Analysis of the domain wall near the top surface}
\label{app:dw}

To obtain the domain wall shape, we numerically solve the variational equation for the domain width $\Delta(z)$~\eqref{eq:dw-width-equation} using the test value $\lambda = \ell$. 
The obtained function is substituted into the expression of the energy density~\eqref{eq:dw-energy-density} and integrated as a function of $\lambda$.
This allows to determine the value of the penetration depth $\lambda$ in the second order and substitute it back into~\eqref{eq:dw-width-equation} to repeat the iteration process until convergence. The relative accuracy of $10^{-3}$ for the domain wall parameters can be obtained within 3--5 iterations. The same procedure is used to analyze skyrmions of small and large radii.

\section{Analysis of the skyrmion shape}
\label{app:sk}

The energy~\eqref{eq:energy} in the cylindrical reference frame $(r,\chi,z)$ reads
\begin{equation}\label{eq:sk-energy-density-gen}
\begin{split}
\mathscr{E}_\text{sk} & = A \left[(\partial_r\theta)^2 + (\partial_z\theta)^2 + \sin^2\theta(\partial_z\phi)^2\right] + K \sin^2\theta \\
& + D \left[\frac{\sin2\theta \sin(\phi-\chi)}{2r} + \sin(\phi-\chi) \partial_r \theta - \sin^2\theta \partial_z \phi\right].
\end{split}
\end{equation}
To analyze skyrmions of small radius, we substitute the Ansatz~\eqref{eq:linear-sk} into~\eqref{eq:sk-energy-density-gen}, which leads to the effective energy density
\begin{equation} 
\mathscr{E}_\text{sk\,sm}^\text{bulk} \approx 38.7A + 2\pi A \left[ \dfrac{(R_\text{sk\,sm}^\text{bulk})^2}{\ell^2} - 4 C \dfrac{D}{D_c} \dfrac{R_\text{sk\,sm}^\text{bulk}}{\ell} \right].
\end{equation}
The condition of the minimum of this expression gives~\eqref{eq:small-sk-radius-bulk}. The 3D Ansatz gives the effective energy density
\begin{equation} 
\begin{split}
\mathscr{E}_\text{sk\,sm} & \approx 38.7A + 2\pi^3 A (\partial_zR_\text{sk\,sm})^2 + 2\pi K R_\text{sk\,sm}^2 \\
& - 2\pi^2 D R_\text{sk\,sm} \cos \left(\frac{\lambda D}{2A} e^{z/\lambda}\right) \\
& - \pi\frac{D^2}{2A} R_\text{sk\,sm}^2 e^{z/\lambda}\left(2 - e^{z/\lambda}\right)
\end{split}
\end{equation}
with the variational equation for $R_\text{sk\,sm}$
\begin{equation} 
\begin{split}
2\pi^2A \partial_{zz}R_\text{sk\,sm} +  \frac{D^2}{2A}  e^{z/\lambda}\left(2 - e^{z/\lambda}\right) R_\text{sk\,sm} \\ = 2KR_\text{sk\,sm} - \pi D \cos \left(\frac{\lambda D}{2A} e^{z/\lambda}\right). 
\end{split}
\end{equation}
and the boundary conditions $\partial_zR_\text{sk\,sm}(0) = 0$, $\partial_zR_\text{sk\,sm}(-\infty) = 0$. 

The effective energy density of the large radius skyrmion in the bulk~\eqref{eq:circular-sk} reads
\begin{equation}\label{eq:large-sk-energy-bulk}
\mathscr{E}_\text{sk\,lar}^\text{bulk} \approx 4\pi A \left[ \frac{R_\text{sk}}{\Delta} + \frac{\Delta}{R_\text{sk}} + \frac{R_\text{sk}\Delta}{\ell^2} - 2C\frac{D}{D_c} \frac{R_\text{sk}}{\ell}\right]
\end{equation}
with the minimum reached at~\eqref{eq:large-sk-radius-bulk}. Taking into account the effect of the surface, the energy density reads
\begin{equation} 
\begin{split}
\mathscr{E}_\text{sk\,lar} & \approx \pi A \left\lbrace \frac{R_\text{sk\,lar}^\text{bulk} \left[ 12 + \pi^2 (\partial_z \Delta)^2\right] }{3\Delta} + 4\frac{\Delta}{R_\text{sk\,lar}^\text{bulk}} \right\rbrace\\
& + 4\pi K \Delta R_\text{sk\,lar}^\text{bulk} - 2\pi^2 D R_\text{sk\,lar}^\text{bulk} \cos \left( \frac{D^2}{2A}\lambda e^{z/\lambda} \right) \\
& - \pi\frac{D^2}{A}\Delta R_\text{sk\,lar}^\text{bulk} \left(2 - e^{z/\lambda}\right).
\end{split}
\end{equation} 
This expression leads to the variational equation
\begin{equation} 
\begin{split}
\frac{\pi^2A}{6\Delta^2} \left[ 2\Delta \partial_{zz}\Delta - (\partial_z\Delta)^2 \right] + \frac{2A}{\Delta^2} \\
+ \frac{D^2}{2A} e^{z/\lambda} (2-e^{z/\sigma}) = 2K + \frac{2A}{(R_\text{sk\,lar}^\text{bulk})^2}.
\end{split}
\end{equation}
with $\partial_z\Delta(0) = 0$ and $\partial_z\Delta(-\infty) = 0$, c.f.~\eqref{eq:dw-width-equation} for a straight domain wall. 

The difference between simulations and analytics is a consequence of the simplified Ansatz~\eqref{eq:linear-sk} and~\eqref{eq:circular-sk}, which does not take into account a fine structure of the radial dependency of $\phi$ and asymptotics for $\theta$ at the origin and infinity. For example, taking into account $z$ dependence of the skyrmion radius in~\eqref{eq:circular-sk} as $R_\text{sk\,lar}^\text{bulk} \to R_\text{sk\,lar}(z)$ in addition to the function $\Delta(z)$, one obtains the boundary condition $\Delta(0)\partial_zR_\text{sk\,lar}(0) + [r-R_\text{sk\,lar}(0)]\partial_z\Delta(0) = 0$. This shows that the condition $\partial_z \Delta(0)$ is not a strict one if a fine structure of the soliton near the surface is taken into account.

\end{document}